\documentclass[a4paper]{article}

\usepackage{graphics}
\usepackage{graphicx}
\usepackage{amsmath}
\usepackage{amsfonts}
\usepackage{subfigure}
\usepackage{longtable}
\usepackage{ccaption}
\usepackage{setspace}

\renewcommand{\vec}[1]{\mathbf{#1}}
\newcommand{\D}{$^{\circ}$}
\newcommand{\wn}{cm$^{-1}$}
 
\newcommand{\A}{\AA{}}

\newcommand{\kbt}{$k_{B}T$}

\title{Self-assembled monolayer molecule dynamics are perturbed by surface- and surrounding monolayer-derived geometrical confinement}
\author{Ella M. Gale\footnote{Affiliation: Department of Chemistry, Imperial College London, London, SW7 4AZ} \footnote{Present Address: School of Experimental Psychology, University of Bristol, 12a, Priory Road,Bristol, BS8 1TU United Kingdom} \footnote{\textit{ella.gale@bristol.ac.uk}}}

\begin{document}
\maketitle

\begin{abstract}
The surface a thin-film is attached to and the surrounding monolayer causes geometrical confinement of a interrogated molecule; we look at the base case of a SC$_{18}H_{37}$ in a SC$_{18}H_{37}$ monolayer on Au[111]. Normal mode analysis was used to get vibrations, and these are analysed using mode character indicators which can quantify: how active an element is in a mode; the overall direction of the mode; and which chemical coordinates are relevant. We examined the 4 possible packing structures. We find that the more thermodynamically stable structures were less perturbed by the surface and more supported by the surrounding monolayer. The surface-perturbed modes were below 100\wn, had a higher global, carbon, sulfur, longitudinal and torsional characters, indicating unit cell backbone motions, often with increased S motion parallel to the surface, and an increased terminal methyl group motion. Modes identified by this technique showed a difference between experimental vibrations (with and without the surface) that was twice as large as those not identified. The surrounding monolayer had a larger effect on a single molecule dynamics than the surface, including stabilising the molecules enough for 12 high energy modes to move $\approx$425\wn down in energy to below $k_B T$, allowing them to be populated at room temperature. These modes had higher local and higher H characters, and were highly modulated by the SAM structure. This work shows novels ways to analyse vibrations, and demonstrates the crucial need to include geometric confinement effects in SAM studies.
\end{abstract}

\section{Introduction}

%This paper will be of interest to surface  scientists, electronic transmission
%theoreticians and modelers, as well as anyone who wishes to use normal mode
%analysis, NMA, on chemical systems.

% Why SAMs? What are they used for? What can we do with them
Self-assembled monolayers (SAMs) have been extensively used in a wealth of  nanotechnology applications, from molecular electronic devices to biosensors  to patterning~\cite{202}. For example, they can be used as coatings to reduce wear and tear on a surface~\cite{247} or to make a surface more hydrophobic~\cite{246}. SAMs are not limited to flat surfaces: coating gold nanoparticles in SAMS has the effect of improved sensitivity in biosensors.~\cite{253} The appeal of self-assembled systems is the possibility of controlling the thin-film properties to a great degree, once their properties are understood. Controlling electronic orbitals is a route to controlling colour, thermoelectric and electric properties.~\cite{257} Electrostatic control can alter the conformation of a molecule, which can effect the frictional force~\cite{265}. 

% confinement
In this paper, we focus on the effect of geometrical confinement,  defined as a physical barrier to a molecule's exploration of space. For molecules embedded in SAMs, this can come from two causes, the surface the SAM is attached to, and the surrounding monolayer.  Generally, the effect of confinement is not included in calculations, with vacuum-minimised or gas phase structures being a common starting point, although there has been a move towards it in recent years.~\cite{266} Also, device properties can be drastically changed under compression, which can be viewed as an extreme form of confinement: compression of two Au-electrodes coated with SAM gave less defects than the bare electrodes~\cite{251} and, compression can result in altering conductance values.~\cite{258}

% vibrations

Similarly, ground state structures are generally used as a starting point for molecular electronic models. As the molecule vibrates, both its structure, as sampled by tunnelling electrons, and its electronic structure, which relates to conductivity and colour, are altered. Thus, we also take into account the non-equilibrium properties introduced by vibration. An interesting recent paper~\cite{254} looked at the heat transfer properties of SAM compared to the bulk. In the bulk state, most of the thermal conductivity is to with intermolecular intermolecular forces like Lennard-Jones interactions and translational motion. For thioalkanes localised on a surface, heat transfer is largely related to the internal coordinate (torsion, angle, bond) motion, with angle and then torsional motion being the most important contributions. The ordering and alignment of molecules on the surface allows the tuning of thermal transfer characteristics via altering microscopic energy transfer modes. As these modes are the vibrations, there is a clear need to understand how these vary and the relationship between structure and vibration energies. 

% why monolayer?
When SAMs are used as the inert matrix in `single molecule' molecular electronics experiments, such as scanning tunneling microscopy (STM), crossed-wire junction and break junctions~\cite{202}, it is usually assumed that the surrounding monolayer has a negligible effect on the conformation of the interrogated molecule. Generally, Pi-Pi interactions are considered when optimising organic electronic devices, however a recent paper~\cite{250} of a mixed ferrocene-thioalkane showed that a more favourable van der Waals interaction between thioalkanes had a large effect on the reproducability, efficiency and yield. This shows the need to consider the surrounding thioalkane monolayer. Experiments show variations in the conductance of molecules embedded to SAMs~\cite{BlinkingExperiment} suggesting that the surrounding SAM may play an indirect but important role. For example, an electro-active molecule in a SAM under an STM tip was observed to stochastically switch from a high to a low conducting state,  which may be a result of vibrations being mediated by the surrounding monolayer.\cite{BlinkingExperiment} 

% why include more of the monolayer?

In this paper, we look at the model system of thioalkane chain SAMs on Au[111] surface. Their appeal has been based on the assumption that, as is typical for alkyl sulfide chains on metal surfaces, they form  well-ordered structures, which can  then provide an inert matrix  for further nanoscale experiments~\cite{243}.  However,  the continuing study of the phase behaviour, structure and dynamics of these materials  reveal a more complex picture~\cite{244}. For example, there has been much controversy in the literature regarding the SAM structure of alkanethiols on gold as well as  the role played by the surface.   To clarify these issues detailed studies have been performed on  short chains (typically up to 8C repeats) as model systems~\cite{242}. However, similarly detailed theoretical work for very long-chain SAMs is currently lacking due to the complexity of the system, even though long chain SAMs are highly utilized experimentally.

To assess the role of the SAM as a constraint to the single-molecule motion, we focus here on the low energy, slow vibrations of long-chain alkanethiol SAMs on gold surfaces. These  modes are both more likely to be occupied, more likely to involve large conformational changes across the whole molecule, and can be used to generate a set of conformers sampled by traversing electrons. We will focus on long chain molecules where these effects may be more pronounced and which are typically used in experiments. The purpose of this paper is to address the question of how a molecule in a self-assembled monolayer is geometrically confined, demonstrate how this confinement alters the dynamics of the system via illustrative examples on several proposed SAM structures, and present a method for how these confinement effects can be included in theoretical descriptions of the system. 
 
The geometrical confinement effects that a molecule embedded in a SAM experiences comes mainly from two sources, the gold surface it is attached to that prevents its exploration perpendicular to the surface, and the surrounding monolayer that confines its motion perpendicular to the monolayer chains (the STM tip is assumed to be sharp enough and far enough away that its confinement effects are small compared to the monolayer and surface). By separating the effects from the two sources of geometrical confinement in the system, we will gain an understanding of the effects of the surface and the monolayer on a molecule inserted in a SAM. 

In this work, we address these effects by doing a normal mode analysis, NMA, of the whole system.  Using a simple classical force-field allows us to investigate long chain thioalkane structures and include the surface and surrounding monolayer. NMA is a useful approach  for getting information about the collective motions of complex systems, and is often used in other soft matter systems, such as studies of protein dynamics~\cite{262} and can help explain experimental results~\cite{263}. 

However, the number of output normal modes is $3N-6$, where $N$ is the number of atoms in the system; which is often an unwieldy number. The vibrations themselves can be difficult to analyse and gain an intuitive understanding. To solve both these problems, we  present qualitative characters that help one gain an intuitive chemical understanding of the system. 

\section{Methodology}

\subsection{Structures of C18 chains on Au(111) self-assembled monolayers}
We focus here on several perfect self-assembled monolayer structures of SC$_{18}$H$_{37}$ chains on Au(111) surfaces as model systems to introduce the methodology. The packing structures are taken from Zhang et al's analysis~\cite{Goddard1} and are denoted here as PSI ($\left(\sqrt{3} \times \sqrt{3} \right)$R30\D), PSII (herringbone), PSIII (zig-zag version $c(4 \times 2)$) and PSIV (diamond-shaped version of $c(4 \times 2)$). The structures are shown in
table~\ref{tab:PSSide}. Throughout the paper, the $x-y$ plane defines the surface, the $z$ axis is normal to the plane and the chains are tiled away from the surface normal by the tilt angle, $\alpha$.

\begin{table}[htbp]	% GRAPHICS: Minimised structures from side
\centering
	\begin{tabular}{|c|c|}
\hline
	\includegraphics[height=148pt,width=212pt]{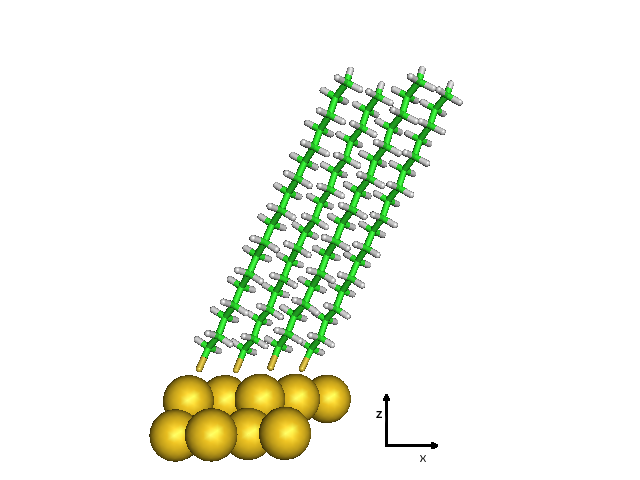}
	&
	\includegraphics[height=148pt,width=212pt]{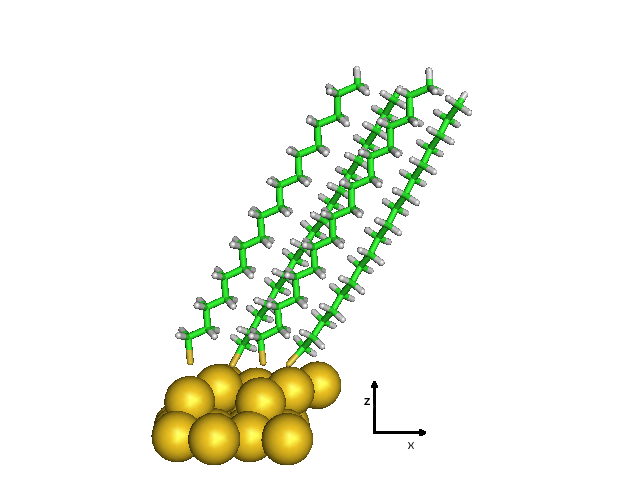}\\
PSI	& PSII	\\
\hline
	\includegraphics[height=148pt,width=212pt]{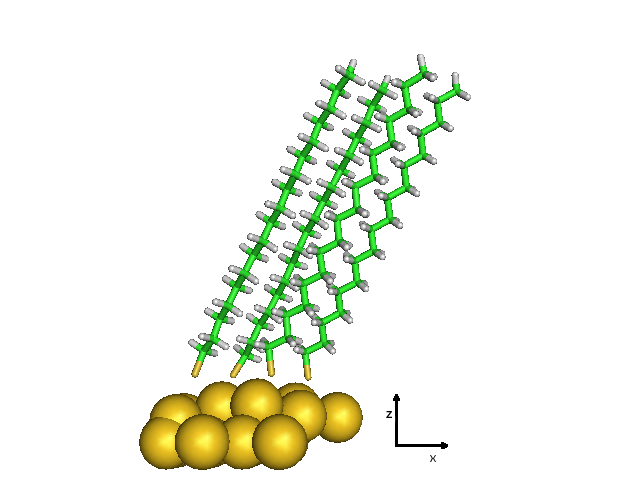}
	&
	\includegraphics[height=148pt,width=212pt]{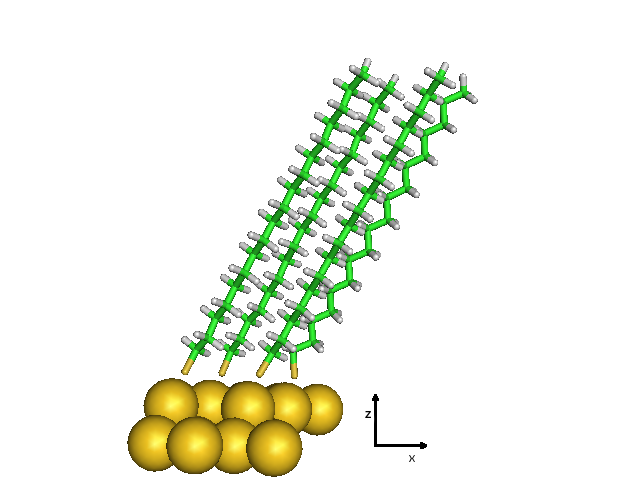}\\
PSIII 	& 	PSIV\\
\hline
	\end{tabular}
	\caption{The minimised packing structures as viewed from the side. In PSI, the rotation angle is the same for all the chains, in PSII and PSIII, two chains of the chains are rotated by 90\D. In PSIV only one chain in rotated. All four structures have the same tilt angle. N.B. the gold surface is the same in all four structures, but has been titled slightly to show the chain orientations better.}
	\label{tab:PSSide}
\end{table}

We use a version of  the Universal Force Field (UFF) which has been 
parameterized to model these structures~\cite{Goddard1}. The UFF is a simple force-field parameterised from small chemical moieties,~\cite{UFF1} not unlike thioalkanes. To simulate the dynamics of this force-field, GROMACs~\cite{Gromacs1,Gromacs2} was modified to include the UFF 3-body angles.~\cite{MyPhDThesis} The gold surface was included only as a constraining barrier, thus the gold atoms were fixed in place, intrasurface chemical coordinates were not input, and the surface only interacted with the chains via the Au-S bond (which was modelled as Morse bond to capture the features of a partially covalent bond~\cite{Goddard1}) and intermolecular interactions (such as Lennard-Jone's potentials). The sulfur adlayer is located in the fcc position, as it had been\ found to be the lowest energy~\cite{Goddard1}.

Values for the rotation angle, $\chi$, and the tilt-angle,  $\alpha$, of the thioalkane chains were identified by minimising the starting structures taken from Zhang et al\cite{Goddard1} over the range $\chi=1-120$\D (those values are due to the 3-fold symmetry of the FCC hollow position) and $\alpha=25-36$\D. Minimisation was done with a steepest descent minimisation algorithm provided within GROMACs. To understand the dynamics, a normal mode analysis, NMA, was performed on the minimised structures in GROMACS and the output modes were analysed, animated and plotted in MatLab.

%NTS Check what chi and alpha are and make sure you got the right one all the
%way through

%The frequencies from the NMA are spread over five frequency ranges, called
%frequency envelopes in this work, and their approximate ranges are: A: 0-700
%cm$^{-1}$; B: 750:1300 cm$^{-1}$; C: 1300-1900 cm$^{-1}$; D: 2800-2920
%cm$^{-1}$; E: 2920-3000 cm$^{-1}$, and there a large gaps where no frequencies
%are found in between them. These are found in all the systems in this paper.

%To create a conformer representative of the structure an electron is likely to encounter,
%each atom was then propagated along the normal mode direction by the amount
%$\Delta g$, given by  $\Delta g_{t} = \sqrt{\frac{k_{B}
%T}{\lambda_{t}}}$~\cite{160}, where $\lambda_{t}$ is the force constant for that
%mode, $k_B$ is the Boltzmann constant, and T the temperature. A 3-21G* Hartree-Fock
%single point energy calculation was performed on the conformers in Gaussian~\ref{Gaussian}, to
%obtain an approximate electronic structure {\bf{what for? do you use this information later? If not remove}.} 

\subsection{Quantifying the surface confinement purturbation on the monolayer}

To elucidate the effect of the surface, the gold atoms were then removed entirely to leave a structure consisting of only the monolayer, referred to as `PSTA' in this work. Without allowing this system to relax, another normal mode analysis was performed (a relaxed version would not be comparable to our thin-film, the unrelaxed version gives information about just the monolayer interactions). The two systems were then compared. As an example, PSI and PSTA are shown in figure~\ref{fig:PSI}(a) and~\ref{fig:PSTA}(b).

To quantify the effect of the surface, we make the assumption that the part of the eigenvector localised on the thioalkane fragment of the structure will be the same in the PS and PSTA datasets, and then look for modes where this is not true to find where the surface has has had an appreciable effect. If the NMA finds similar vibrational
modes for the two structures, which we would expect as the structures are similar, the eigenvectors should be close to parallel (note, we are comparing eigenvectors from two different structures, eigenvectors in the same structure are all 90\D by definition). To test our assumption, we take the part of the PS eigenvectors located on the thioalkane moities and compute the overlap between them and the eigenvectors from the PSTA normal mode analysis. The overlap between the eigenvectors is the cosine of the angle between them and a value of 1 indicates that the two eigenvectors are parallel,
i.e. identical (within numerical accuracy).~\footnote{To be clear, the longitudinal character computes the angle between the carbon backbone and mode to measure how much of the mode is along the carbon backbone, and here we are computing the angle between two normal modes in different runs to measure similarity in normal mode space.} We compare the eigenvectors between the two sets in order of energy. The assumption that the eigenvectors are similar with and without the surface was found to be approximately correct, as we could be stringent with the numerical cut-off and designate modes with an overlap that did not equal 1 to three decimal places (the coordinates of the atoms were only known to this precision) as having been affected by the presence of the surface. 

The eigenvectors are ordered by energy and the lowest energy 72 modes from the PS datasets are discarded to remove modes associated with the gold-gold bonds; the gold atoms were used as an unmoving confining surface and Au-Au interactions were not given in the model. However, by assuming that the first 72 modes are unphysical we could perhaps be throwing away valid modes that are anomalously low in energy. For this reason a second analysis was done where the subset of PS modes were picked from the total; 72 modes were discarded, but these were not necessarily the lowest energy modes. To do this the assumption was made that the modes hadn't changed in form, but only order/energy. For this reason the 672 modes that had the highest overlap with the PSTA modes were kept. The results from this analysis agreed with those from the analysis presented here, and show that our first assumption was not a bad one. Watching animations of some of the first few modes confirms that they are intrasurface vibrations Further details on this second analysis can be found in reference~\cite{MyPhDThesis}.

\subsection{Quantifying the surrounding monolayer purturbation on a single thioalkane molecule}

To elucidate the effect of the surrounding monolayer on the ``single-molecule" chain, a thioalkane fragment
SC$_{18}$H$_{37}$, denoted `C18', was minimised \textit{in vacuo} using GROMACS and then compared to the  PSTA structures. The thioalkane fragment is shown in figure~\ref{fig:C18}(c).

\begin{figure}[!tbp]
\centering
\subfigure[]{\includegraphics[width=0.32\textwidth]{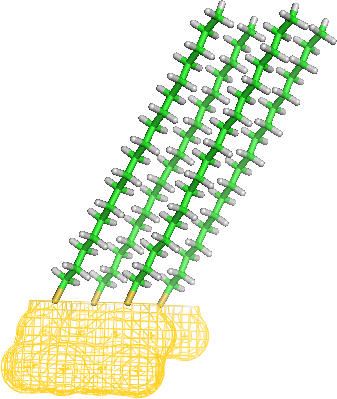}} 
\subfigure[]{\includegraphics[width=0.32\textwidth]{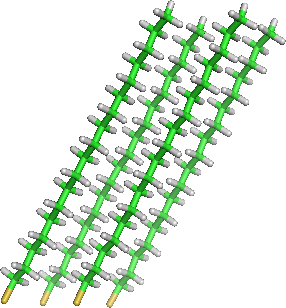}}~
\subfigure[]{\includegraphics[height=4cm]{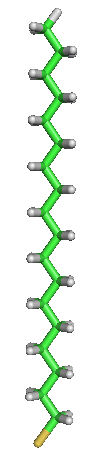}}

\caption{The different structures used to elucidate the effect of
 the surface and the monolayer on the vibrational modes. The PSTA
structures, b, are created by removing the surface from the SAM
minimised structure, a. The C18 structure, c, is
minimised \textit{in vacuo}.}
\label{fig:GeneralStruct}
\end{figure}

\subsection{Character classification  of the normal modes}

In order to extract useful information from the high-dimensional results, we followed a post-analysis often used in the field of protein dynamics. Karplus, Brooks et al~\cite{62} formulated a qualitative test to determine how global a protein normal mode is by summing the normal mode atomic displacements over all degrees of freedom, $k$, to give two character indicators: the global character indicator and the local character indicator. A mode is considered highly globalized if the global character indicator is high and the local character indicator is low. Similarly, a mode is localised if the local character is high and the global character is low. 

For clarity, the global and local characters from~\cite{62} are reproduced below.

$$
\mathrm{Local \; character} = \sum\limits_{k-1}^{3N} y_{ki}^4 \; . $$
$$
\mathrm{Global \; character} = \left( \frac{\sum\limits_{k-1}^{3N} |y_{ki}| }{\sqrt{3N}}\right)^{4} \; .
$$

As all normal modes are normalised to 1, a global mode will have many, low value components indicating many coordinates are taking part in the vibration. In the limit of a perfectly global mode they would all be equal (this would correspond to a translation), which gives the local character the minimum value of $(3N)^{-1}$ and the global character its maximum value of $9N^2$. Contrarily, a localised mode would have very few, high value components, and the most local mode would be a single component with the value 1, all other components being zero; this gives the local character its maximum value of 1, and the global character its minimum value of 1. Most modes will reside between the maximum and minimum possible character values, and so we must compare the values of the two indicators to qualitatively classify their behaviour.

Following this example, we produced a series of  character indicators in order to classify the normal mode dynamics of the SAMs. To identify which types of atoms within the system are moving in a given normal mode, $i$, the elemental character indicator is
used, as given by:

$$
\mathrm{Elemental \: character }\; = \; \frac{ \sum\limits_{k,\, k \in E} | y_{ki} | }{
\sum\limits_{k =1}^{N} | y_{ki} | } , 
$$

where $E$ is the set of atom indices that correspond to the element we are interested in,  $\vec{y}$ is the normal mode column vector from the Hessian matrix and $j$ is the degree of freedom. The sum in the numerator is taken only over the degrees of freedom that correspond to the element under investigation. The absolute value of the displacement is used to avoid excluding components of the eigenvector that are anti-parallel to each other. The denominator is the sum of the absolute maximum displacements for all components on all atoms of that eigenvector. The elemental character is useful for spotting motions of the carbon backbone or sulfur adlayer.

The Cartesian character indicator is used to classify the direction of the normal mode displacements. This is useful as the gold surface was placed on the $x$-$y$ plane, so motions perpendicular to that plane have a high $z$-character, and those parallel to the plane would have a low $z$-character, high $x$- or $y$- characters. The Cartesian character is calculated similarly to the elemental character:  the absolute value of the maximum displacement for either the $x$, $y$ or $z$ components of the eigenvector on each atom is summed, and measured as a proportion of the total sum of components, as given by: 
$$
\mathrm{Cartesian \: Character } \; = \; \frac{ \sum\limits_{k, \, k \in X} | y_{ki} | }{
\sum\limits_{k = 1}^{N} | y_{ki} | }, 
$$

where $X$ is the set of eigenvector components of a particular Cartesian direction. 

The eigenvector $\vec{y}$ is in Cartesian difference coordinates, $\Delta\vec{
x}$, and can be converted to internal chemical coordinates, $g$, using the Wilson B Matrix, $\vec{B}$ as $\vec{g} = \vec{B} \Delta\vec{x}$.~\cite{WDandC}.  Thus, we can obtain a chemical coordinate character indicator as

$$
\mathrm{Chemical \: Coordinate \: Character } \; = \; \frac{ \sum\limits_{k=g_{1}}^{g_{n_{g}}} |\vec{B}\vec{y_{i}}(k)| }{ \sum\limits_{k =1}^{N_{D}} |
\vec{B}\vec{y_{i}}(k) | } , 
$$

where $n_{g}$ is the number of chemical coordinates of type $g$ and $N_{g}$ is the total number of chemical coordinates. The chemical coordinate character indicator can be used to identify whether the motion is mostly torsional, for example.

To measure how much of the motion is parallel to the carbon chain the longitudinal character indicator can be used. This character indicator is specifically for all-trans alkanes. The order parameter for chain $A$, $\vec{R}_{A}$,~\cite{212} gives the direction of the carbon backbone and this is calculated and normalised. The longitudinal character indicator is calculated by comparing the overlap of the atomic displacement vector with the carbon back bone vector. Overlap is calculated from the angle between backbone vectors using the dot product. The longitudinal character is given by:

$$
\mathrm{Longitudinal \: character }  \; = \; \sum\limits_{A=1}^{C}
\sum\limits_{n=1}^{N} \frac{1}{q} \frac{\vec{y}_{i}(n) \cdot \vec{R}_{A} }{
\vert \vec{y}_{i}(n) \vert \vert \vec{R}_{A} \vert} .
\label{eqn:LongChar}
$$

The sum is over the number of chains in the system, $C$. The sum excludes the components which are less than the minimum significant overlap value, $q$, which is any normal mode component whose modulus value is less than $\frac{1}{2}\frac{1}{3N}$.

\section{Results}

\subsection{The structure energies}

The energy relations for PSI-PSIV matched the Zhang's energies~\cite{Goddard1} to within -9.3kcalmol$^{-1}$ for PSI, +1.1kcalmol$^{-1}$ for PSII and 0.2kcalmol$^{-1}$ for PSIV, where PSIII is taken as the zero point in both cases. The discrepancy is likely due to the use of a different minimisation algorithm. For all structures, there was a relaxing of the sulfur adlayer away from hexagonal symmetry and $\alpha$, the tilt angle, varied slightly between chains in the unit cell. This relaxing of the sulfur adlayer shows how important the Au-S bonds are in modeling these systems. Other simulations~\cite{SciencePaper08} have shown the gold surface relaxing and distorting as well as the adlayer, but in this work the gold surface was not allowed to relax, although, due to the mass of gold, this introduces only a small error.

The frequencies from the NMA are spread over five frequency ranges, termed frequency envelopes, and their approximate ranges are:
A: 0-700\wn; B: 750-1300\wn; C: 1300-1900\wn; D: 2800-2920\wn; E: 2920-3000\wn.  There are relatively large gaps between them where no frequencies are found. These frequency envelope structures are found in the NMA for all the structures in this paper and result from the values of the fundamental frequencies associated with the chemical coordinates of the system. Since we are interested in the low energy modes, we will mainly concentrate on modes within envelopes A and B.

\begin{table} %	 PSNRG Minimised energies
\centering
 \begin{tabular}{|c|c|c|c|c|c|}
  \hline
	Packing 	&Energy & \multicolumn{4}{c|}{Values of beta}\\
	Structure 	& 	&Chain 1 & Chain 2 & Chain 3 & Chain 4\\
		  	& kcalmol$^{-1}$ & \D & \D & \D	     & \D	\\
\hline
	PSI	& 14.004 & 120	& 120	& 120	& 120	\\
	PSII	& 4.374  & 120	& 120	& 150	& 150	\\
	PSIII	& 0.0	 & 150	& 120	& 120	& 150	\\
	PSIV	& 0.465  & 150	& 120	& 120	& 120	\\
\hline
 \end{tabular}
\caption{The relative energies for the four packing structures after minimisation. The monolayers were minimised with the chain rotation angle $\beta = 90$\D ($\beta$ is the angle of the carbon backbone relative to one another), the rotation angle (of carbon backbone relative to the surface) $\chi = 48$\D, and the tilt angle (of the carbon backbone relative to the surface normal), $\alpha = 32$\D. This was the only data set with the energy of packing structure IV less than 1 kcalmol$^{-1}$ above the energy for
PSIII.}
\label{tab:PSNRG}
\end{table}

\subsection{The effect of the surface}

We found that  the  effect of the surface varied according to the
packing structure of the monolayer, namely PSI and PSII had 36\% and 44\% of the modes with non-unity overlap, whereas the more stable structures, PSIII and PSIV had only 20\% and 26\% modes affected.  Additionally, most of the low energy modes (first 100 modes,
$<$300\wn) are affected regardless of structure.  Within the
A envelope very few modes are affected by the surface for the more stable structures III and IV. The less stable structures I and II have more surface-affected modes above this energy.  

To quantify these effects further, the gold is removed from the PS data sets so that the PS and PSTA  are the same size, and it is possible to examine the change in energy between the congruent modes in each set; this quantity was designated $\Delta \omega_{PSTA\rightarrow PS}^{EV}$, where the `EV' superscript designates that the modes are ordered by their index, and thus by energy. $\Delta \omega_{PSTA\rightarrow PS}^{EV}$ is defined as $ \Delta \omega_{PSTA\rightarrow PS}^{EV}(i) = \omega_{PS}(i+72) - \omega_{PSTA}(i) ,\: i = 1,...,i=672. $

$\Delta \omega_{PSTA\rightarrow PS}^{EV}$ is not large, individually the effect is biggest for the low energy modes, with a large percentage change for modes below 100\wn. The packing structure modulates the surface effect; PSI shows a curve indicating that that the addition of the surface causes these modes to be raised in energy. For the other three structures the effect is not as simple, and $\Delta \omega_{PSTA\rightarrow PS}^{EV}$ oscillates around a trend like that seen for PSI, which suggests that although the low energy modes are raised in energy overall, the modes are being reordered as well, and could suggest whole unit cell vibrations which are modulated by the unit cell structure.

The characters were averaged and compared between the two subsets to identify the types of modes that are most sensitive to the surface. 
Compared to the unaffected modes, the surface-affected modes had a higher global,longitudinal and carbon characters, plus lower local and hydrogen characters. There was a slight increase in bond and torsional characters and a slight decrease in angular characters. It is important to note that there was no correlation between the size of the overlap and any of the characters, despite some characters having been designed to pick responses to the surface (Cartesian character), and the monolayer (Longitudinal character), demonstrating that the dynamics are not simple.

The high global and carbon characters in surface-affected modes implies backbone modes (modes where the entire carbon backbone deforms). For these modes the sulfur character was also much higher, as these are the modes with a large magnitude of vibration at the surface, in agreement with recent work~\cite{SciencePaper08}. Low $z$ character and raised $y$ character, taken with an increased longitudinal character, imply  longitudinal modes. The presence of
both low angular character and lowered Hydrogen character implies backbone waving type motions rather than internal backbone concertinas because concertina modes are vibrations of three body bond angles. The increased bond character but decreased Hydrogen character shows that the C-C and C-S bonds, which form the backbone of the molecule, are being distorted. The backbone modes are those with almost equal H and C character; the H-C based modes are those with a high H and lower C character because a hydrogen will move more than a carbon in a vibration of a C-H bond. Thus, we see that the surface has the biggest effect on global backbone motions, especially those with a large longitudinal component, because the surface constrains longitudinal motion. A recent paper~\cite{267} performed a vibrational analysis of an azobenzene SAM which can take both cis and trans forms, it was found that the trans form had more of its vibrational components perpendicular to the surface resulting in a higher stiffness of the monolayer, this suggests that the geometrical confinement of the surface could alter SAM stiffness.

\begin{figure}	% Overlap PSTA\rightarrowPS (not sorted by overlap but compared
by index)
\centering
 \includegraphics[width=13.54cm,height=10.17cm]{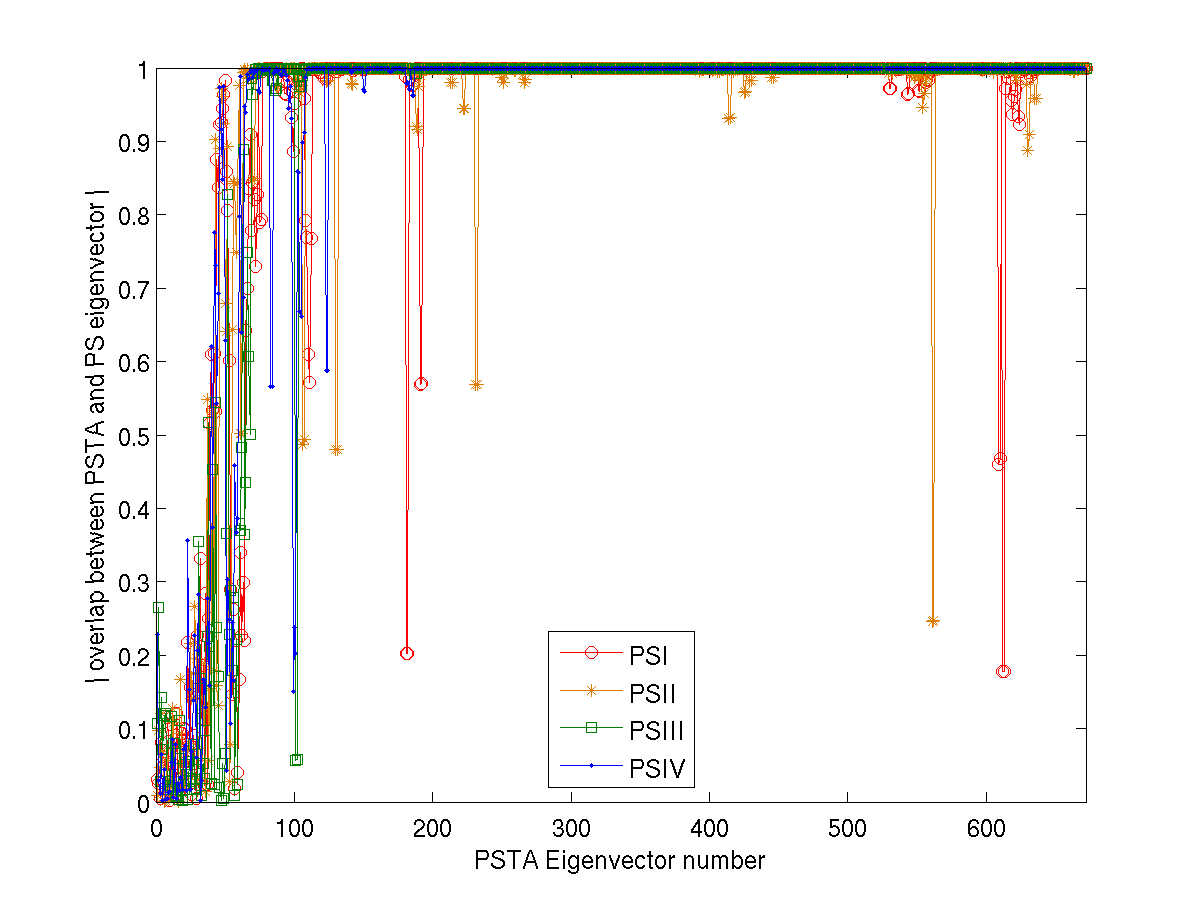}
 % OverlapPSTAPSAll.png: 1200x901 pixel, 150dpi, 20.32x15.26 cm, bb=0 0 576 432
 \caption{The overlap between PSTA and PS normal modes defined as the angle between the two vectors in Cartesian coordinate space. A value of 1 or -1 means the vectors are parallel or anti-parallel respectively. A value of zero means the vectors are perpendicular. This graph shows the effect of the gold surface on the system.
 These are mostly localised to the first 100 modes
($\sim$1-$\sim$200\wn), but there are outliers all along.}
 \label{fig:OverlapPSTAPSAll}
\end{figure}

By plotting the overlap between the PSTA and PS datasets against the mode energy, we can identify exactly which modes are affected by the surface and the degree to which they are affected (figure~\ref{fig:OverlapPSTAPSAll}). The plot shows a general S-shaped structure spread over the low energy modes (0-120\wn), with troughs of highly affected modes superimposed on it. Above this energy range, the overlap is close to one, with just a few highly surface-affected modes. Inspection of the animated vibrations for these modes confirms the results from the average character analysis described above, as well as showing any differences within the subset of affected modes, such as the high energy modes in envelopes D and E, which are global C-H bond movements. Most affected modes appear to have some movement on the sulfur; the largest sulfur movements are on the low energy modes or on the carbon bonded to it which is a higher energy.  For the low energy modes ($<$150\wn), all four structures have troughs. Above this energy, only PSI and
PSII have large troughs, and so are affected at these energies, demonstrating that the more thermodynamically stable packing structures are less perturbed by the surface.

The troughs between 93-140\wn (mode no. 100-150) are all carbon backbone modes (both longitudinal and lateral) with a deformation of the sulfur adlayer parallel to the surface that leads to a
deformation of the internal volume in the unit cell. The troughs in the range 226-239\wn (mode no 170-180) are carbon backbone motions (longitudinal and lateral) with sulfur adlayer motion perpendicular to the surface. 

Similar modes appeared at similar energies across the structures. This is a conceptual separation of the modes expected from the
connectivity of the system and the actual differences in the structures. As an example, for the 2$^{\mathrm{nd}}$ peak, at $\sim$ 230\wn, all the packing structures have a peak 2 EV wide, whose animated vibrations look similar, and has been assumed to be
the same peak in the different structures. This peak appears at slightly different energies across the different packing structures. This implies that a simple model of a $\sqrt{3} \times \sqrt{3}$ packing structure will give some indication of where surface affected modes will be, but including the correct packing structure will alter their energy. However this only holds over for an energy below 250\wn, as some modes identified by an analysis of PSI are not seen in PSIII and PSIV, demonstrating that stable structures can stabilise the monolayer vibrations against perturbation by the surface.

\begin{figure}
 \centering
 \includegraphics[width=13.54cm,height=10.17cm]{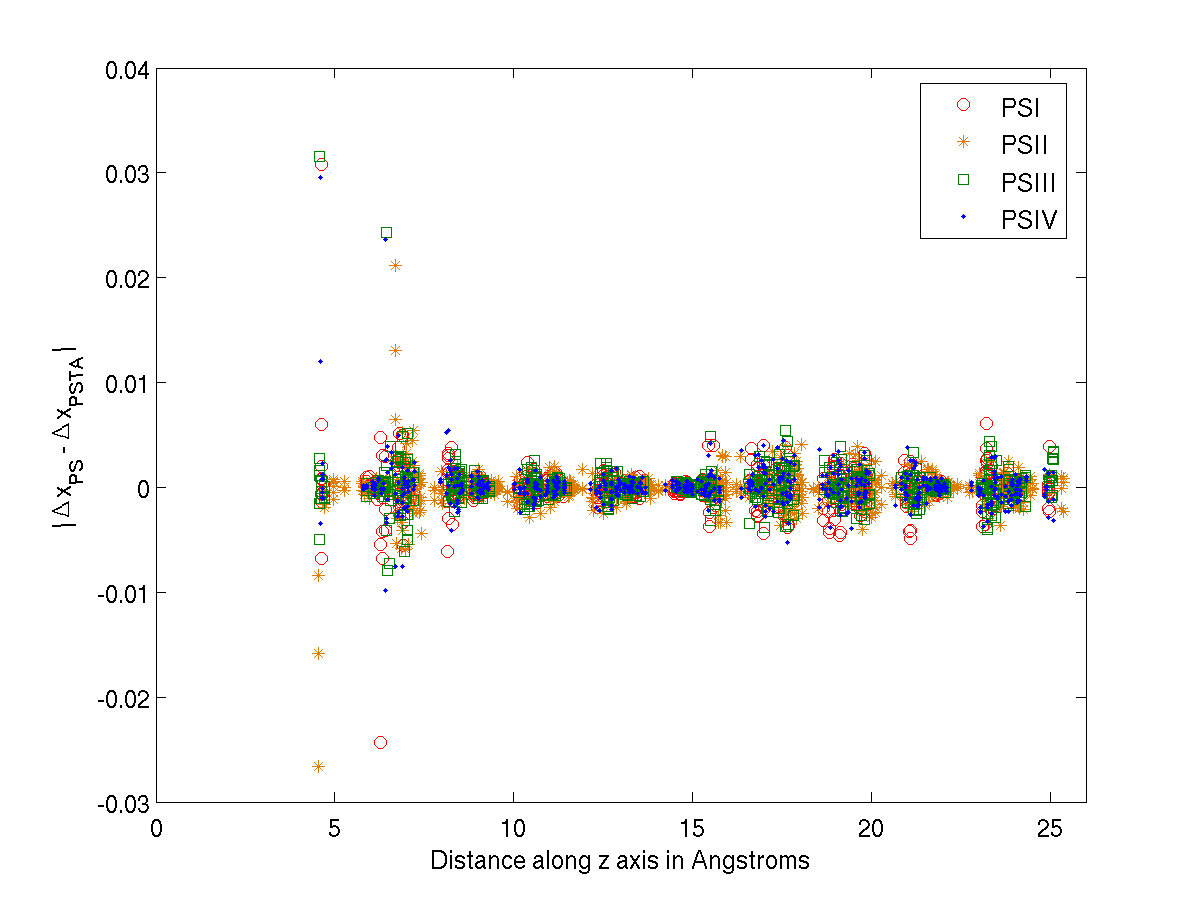}
 % DeltaDeltaXPSTA_PSAll.png: 1200x901 pixel, 150dpi, 20.32x15.26 cm, bb=0 0 576
 \caption{The change in the components of eigenvectors, $\Delta x$, plotted against the distance from the edge of the box. The gold surface is at around 4 \A. The first line of points on the left are the sulfurs, then the data is clustered into C$_2$H$_4$ blocks. All PS show are large effect at the sulfur and on the 1$^{\mathrm{st}}$
two carbons.}
 \label{fig:DeltaDeltaXPSALL}
\end{figure}

Modes from the two datasets can be compared on a degree of freedom by degree of freedom basis to identify which parts of the chain are causing the modes to be different, or to put it another way, which parts of the molecule are being affected by the surface enough to cause the entire eigenvector to differ. It could be possible that the entire chain is affected and the modes are not similar at all, but analysis of the mode animations showed that even the modes that
were identified as having a low overlap were similar in form. For all packing structures, the largest effect was found to be on the sulfurs; this is not surprising as the sulfurs are closest to the gold surface. An example of a typical eigenvector is shown in figure
~\ref{fig:DeltaDeltaXPSALL}. The 4$^{\mathrm{th}}$ trough at $\sim$ 892\wn has a high degree of motion on the CH$_3$ end of the chain, and is also highly altered. As the sulfur end of the chain is being constrained, the other end to vibrate more freely, which is why modes with a high degree of motion localised on the CH$_3$ end of the chain are affected by the surface. Given that the interface thermal conductance is crucial for understanding nanoscale thermoelectric devices~\cite{264}, surface-caused geometrical confinement from the A-S end of the molecule can alter the phonons at the terminal methyl group, and thus a SAM-devices thermoelectric properties; a result that is not obvious.

Because the force-field used in this analysis includes no quantum
mechanical effects, the mode energies are not directly quantitatively comparable to the real world experimental data; nonetheless it possible to compare the data qualitatively to IR and Raman selected vibrations, especially as the force-field was parameterised from similar experimental data. As a further test of the approach presented here, in table~\ref{tab:Exp}, we compare IR and Raman active vibrations from the two closest experimental versions of the conceptual subsets, namely SAMs on gold and SAM liquid films\footnote{Note, the liquid film is not directly
comparable to the PSTA results as the liquid film is in a stable state, whereas the PSTA is not re-minimised after the removal of the gold in order that the only difference between the PS and PSTA structures is the presence or absence of the surface. This is an example of when a computer simulation allows the performance of experiments not possible in the real world.} and matched to
simulated vibrational results (see Table\ref{tab:Exp}). Modes identified in the simulation to be highly affected by the surface had, on average, 30.6\wn difference between the SAM and corresponding liquid film, whereas those that lacked a large overlap in the simulation data had, on average, a difference between 15.9\wn and 12.8\wn (there are two values as there are two possibilities for matching energy levels and the averages correspond to the
smallest and largest answers), demonstrating that this method can identify modes which are sensitive to the surface, and, as only mechanical effects were included, this demonstrates that part of effect of the surface on SAMs is due to a geometric frustration of explorable space (rather than the surface's electronic levels) and that such effects can be included in a calculation using a cheaper level of theory.

\begin{table}[htbp]
`\begin{tabular}{|r|r|r|r|}
 \hline
 Vibration on	& Vibration in 	& Values picked out & Difference	\\
 gold 		& liquid  	& as being	   & between\\
 surface$^{a}$	& film$^{b}$ 	& heavily affected & experiments \\
 		&		& by the surface$^{c}$	&		\\
 \wn		& \wn		& \wn			& \wn		\\
 \hline
 641		& 731		& -			& +90		\\
 706		& 707		& -			& +1		\\
 715		& 731		& - 			& +16		\\
 720		& 720		& -			& 0		\\
 766		& 731		& 892			& -35		\\
 925		& - 		& 909			& -		\\
 1050		& -		& 1059 			& -		\\
 1230		& 1221 or 1240	& -			& -9 or +10	\\
 1265		& 1258 or 1276	& -			& -7 or +11	\\
 1283		& 1276 or 1294  & -			& -7 or +11	\\
 1330		& 1312		& -			& -11		\\
 1455		& 1474		& 1507			& +19	 	\\
 2854		& 2850		& -			& -4		\\
 2860		& -		& -			& -		\\
 2880		& 2918		& 2880			& +38		\\
 2907		& 2918		& -			& +11		\\
 2925		& -		& 2947 or 2948		& -		\\
 \hline
\end{tabular}
\caption{The IR and Raman active vibrations of thioalkane self-assembled monolayers on a gold surface are compared with those measured in a liquid film. Where possible, using the proximity in energy and the assignments given as a guide, the two sets were compared and a difference in energy found. It was found that on average, the modes which were picked out as having been affected by the surface showed a larger change in energy between the two experiments than those not picked out. $a$ taken from ~\cite{SocratesIR} and the Spectral Database for Organic Compounds, published on the internet by the National Institute of Advanced Industrial Science and Technology (AIST), Japan. $b$ taken from~\cite{239} and ~\cite{240} and references within}
\label{tab:Exp}
\end{table}

%\begin{figure}	% PSTA\rightarrow PS All PS
% \centering
%
%\includegraphics[width=13.54cm,height=10.17cm]{DeltOmPSPSTAAllSmallerRAnge.png}
 % DeltOmPSPSTAAllSmallerRAnge.png: 1200x901 pixel, 150dpi, 20.32x15.26 cm, bb=0
%0 576 432
% \caption{There is a difference between the PSTA and PS data sets. The
%differences are relatively minor and localised to the first 100 PSTA modes, ~
%200\wn. The difference between the two data sets is different for the different
%structures. PSI shows a decaying difference. The other three data sets are much
%more oscillatory. This is just for the energies, this does not take into
%the overlaps.}
% \label{fig:DeltOmPSPSTAAllSmallerRange}
%\end{figure}

\subsection{The effect of the monolayer}

To quantify the effect of the monolayer, we compare 4 non-interacting C18 chains with the PSTA  model. In this way, the interchain interactions in the monolayer model can be assessed.
%The vibrational levels for four non-interacting chains would simply be four
%versions of the vibrational energy levels for one chain, and therefore to calculate the
%frequency density for four non-interacting chains, we multiply the frequency
%density for one chain by four.  By comparing this with the frequency density for
%the PSTA yielded from calculations the effect of including interchain
%interactions in the monolayer model can be assessed. 
%The PSTA frequency density is not badly
%approximated by the non-interacting chain structure frequency density, however 
We find that the 
%overestimates the size of the larger peaks in the PSTA spectra and is also
% at the frequency envelope edges. 
standard deviation between each of the PSTA and 4 x C18 frequency densities are 1.107, 1.120, 1.136 and 1.080 for PSI, PSII, PSIII and PSIV respectively.  Interestingly, structures III and IV are  different in this regard, with interchain interactions having a much larger effect on structure III than on structure IV. As these structures are so close in energy, this result could indicate that interchain interactions dominate the interconversion between the
two structures. 

Both C18 and the PS frequencies show the frequency envelope vibrational structure as described previously. We expect that a complex structure's dynamics (such as PSTA) would be based on the dynamics of a simpler structure (such as C18) that makes up its component parts, because the same chemical coordinates are present.
%(for an analysis of where the frequency
%envelope structure comes from, see~\cite{MyPhDThesis}). 
For this reason the mode frequency densities of each envelope can be predicted from the non-interacting chain model and any discrepancies would be due to interchain interactions. We find twelve fewer modes in the B envelope than expected and twelve extra in the A envelope. 

\begin{figure}[ht]
\centering
\subfigure[A close up of the A-envelope showing the A sub-envelopes.]{
 \includegraphics[width=13.54cm,height=10.17cm]{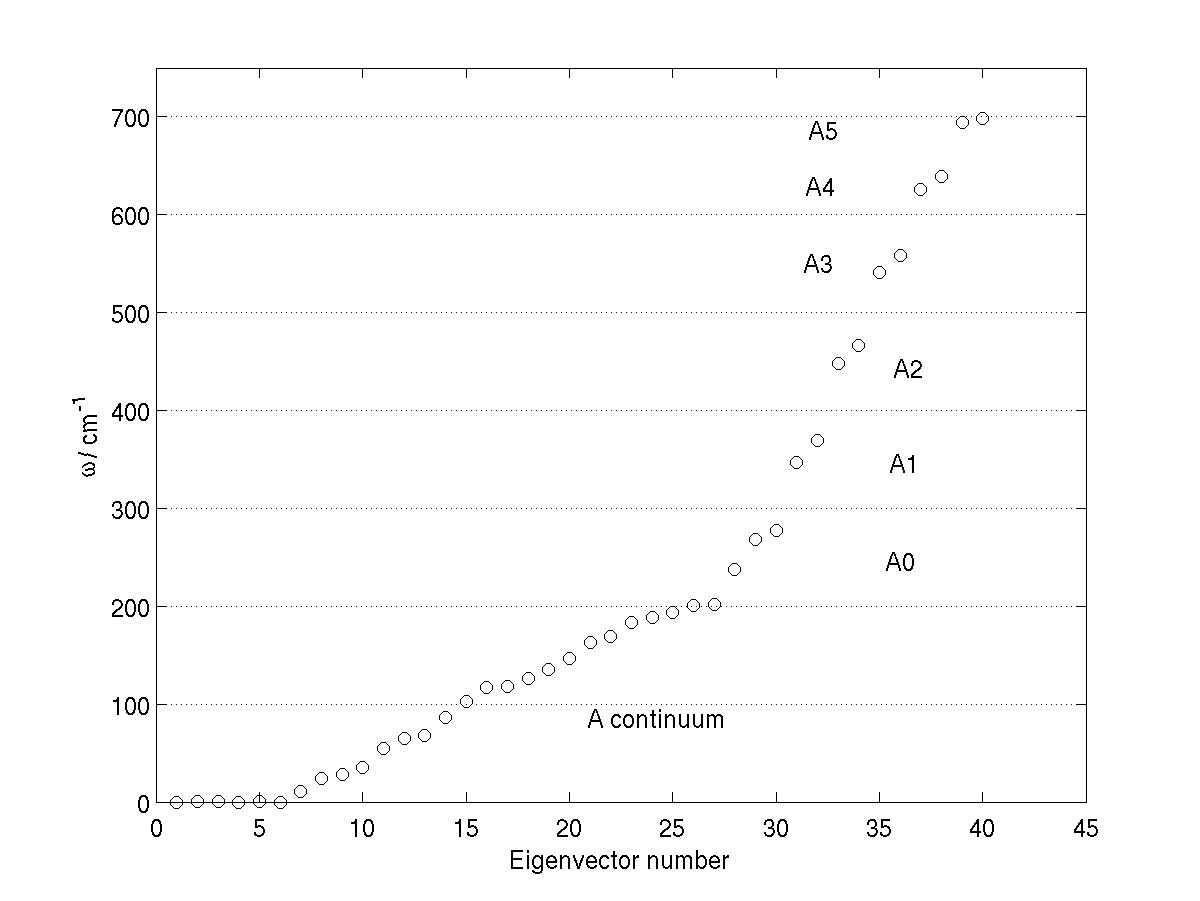}
 % OmC18A.png: 1200x901 pixel, 150dpi, 20.32x15.26 cm, bb=
 \label{fig:OmC18A}
}
\subfigure[A close up of the A region of the normal mode analysis spectra,
showing the different structures of the sub-envelopes.]{
 \includegraphics[width=13.54cm,height=10.17cm]{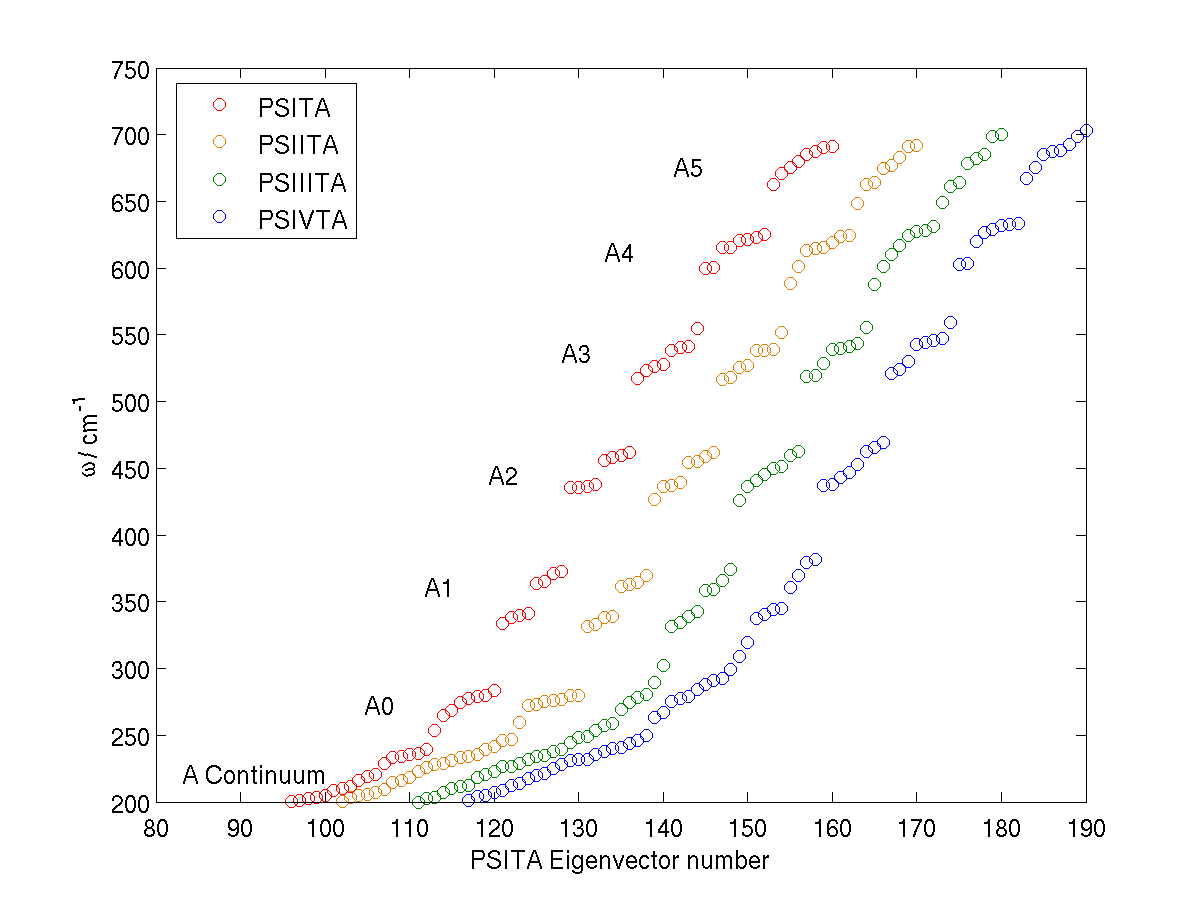}
 % OmC18A.png: 1200x901 pixel, 150dpi, 20.32x15.26 cm, bb=
 \label{fig:OmPSTAllA}
}
\caption{A comparison of the A frequency envelope sub-structure for C18 (a) and the same region for the four packing structures (b). The inclusion of the monolayer affects has substantially affected the the A envelope sub-structure by moving 12 modes from the B envelope down to the A$_{continuum}$. The degree and nature of this effect is modified by the packing structure.}
\label{fig:Subenv}
\end{figure}

The A envelope can itself be viewed as a collection of sub-envelopes, numbered from A$_0$ to A$_5$, which cover the energy ranges from \~ 200\wn to 700\wn; below 200\wn there is no
visually discernible structure and the modes are close together in energy in a sub-envelope called A$_{continuum}$; this sub-envelope structure is shown for `C18' and `PSTA' structures in figure~\ref{fig:Subenv}. We can identify where the extra modes are
added to the A envelope by comparing the number of modes in each
sub-envelope with the number expected from the non-interacting models, and we see that  most of the modes are added to A$_{continuum}$. There is a slight difference in exactly where the modes are added for the different packing structures; PSI has only 10 modes added to A$_{continuum}$, with the extra two modes going into A$_0$. PSIV has 12 while  PSIII and PSIV have 16 and 18 modes respectively added to A$_{continuum}$, with the extra modes come from A$_0$. From examining figure~\ref{fig:OmPSTAllA} we can see
that the extra modes extend the upper frequency of the `continuum' so there is no real gap between A$_{continuum}$ and A$_0$, and for PSIII and PSIV there is shardly any gap between A$_0$ and A$_5$. There is also a modulating effect due to the monolayer structure at work here, see table~\ref{tab:ASubExtraModes}, which shows exactly where the modes end up for each structure.

This  demonstrates that including the surrounding
monolayer moves some high energy modes from the B envelope down in energy by
approximately
%\footnote{We don't know which modes
%have been affected by the monolayer or where they are in A$_{continuum}$, but if
%we assume that they were the lowest in energy in the B envelope,
%$\omega=750$\wn, and were moved to the top of the new A$_{continuum}$/$A_0$
%range, $\approx300$\wn for PSI, $\approx 325$\wn for PSIV (see
%figure~\ref{fig:OmPSTAllA}), then we get a lower bound on the change in energy
%of 425\wn, it could be higher.} 
425\wn. Crucially, modes that were significantly higher than \kbt for the chain \textit{in vacuo}, and would thus not be included in any analysis of molecular vibrations, are now below \kbt and are especially relevant as they would be populated at room temperature.

\begin{table}[htpb]
 \centering
 \begin{tabular}{|c|c|c|}
\hline
  PS		& A-continuum	& A$_{0}$	\\
\hline
 PSITA		& +10		& +2		\\
 PSIITA		& +16		& -4		\\
 PSIIITA	& +18		& -6		\\
 PSIVTA		& +12		&  0		\\
\hline
 \end{tabular}
\caption{Where the 12 extra modes from envelope B end up. For PSIITA and PSIIITA some of the modes expected in A$_{0}$ end up in A-continuum.}
\label{tab:ASubExtraModes}
\end{table}

If the more complex PSTA modes are based on the C18 modes, we would expect the mean characters to be the same over the same energy range of the spectra, and any difference should be due to the inclusion of the monolayer. A$_1$ to A$_5$ are grouped together due to their similarity and because they do not gain or lose any modes as a result of the monolayer. If the character for a region differs from the value for the rest of the normal modes, it could be due to the affect of the monolayer on A or, as it is known that the modes in different parts of the spectra have different characters, due to the difference between the envelope A and the rest of the modes. To
separate out these two effects, we look at the change in percentage difference in character between C18 and PSTA, $\Delta_{C18 \rightarrow PSTA} \Delta_{A \rightarrow \neg A}$, which is given by
$$
\Delta_{C18 \rightarrow PSTA} \Delta_{A \rightarrow \neg A} = 
\frac{A_{n}^{TA} - \neg A^{TA} }{\neg A^{TA}} - 
\frac{A_{n}^{C18} - \neg A^{C18} }{\neg A^{C18}}
\times 100 ,
$$

where the percentage difference in character between the A envelope and the rest of the vibrational spectra, (not-A or $\neg A$), is given by

$$
\Delta_{A \rightarrow \neg A} = \frac{A_{n}^{TA} - \neg A^{TA} }{\neg A^{TA}}
\times 100
$$

where $A_{n}^{TA}$ is one or several of the A sub-envelopes in the PSTA spectra, $\neg A^{TA}$ is the rest of the spectrum that is not in the A envelope. Because the PSTA system is larger than C18 system, the characters are divided between more atoms and may then be lower than expected. For this reason, the difference in character between envelope A and the rest of the spectra is compared with the
difference expected from the C18 data set. $\Delta_{C18 \rightarrow PSTA} \Delta_{A \rightarrow \neg A}$ reports on whether the change in character is more than expected and the sign of this quantity indicates whether it is a change in the same direction as the C18 case. $\Delta_{A \rightarrow \neg A}$ indicates how big the change actually is. 

The global character increases in the A-continuum envelope by about 93.2\% when compared to the rest of the spectra, and the local character decreases in the A-continuum by about 33.6\%. In the C18 data set the A-continuum envelope has a lower global and higher local character than the rest of the non-A part of the spectra, so the change in the PSTA is in the opposite direction to the expected
change. This could be due to the form of the modes being affected by the monolayer or it could be entirely due to the gain of modes from the B envelope.

To elucidate where the monolayer has had an appreciable effect, we can look for the PSTA modes that have the maximum overlap with the C18 root modes. Unlike the analysis of the affect of the surface, the index of the PSTA modes from the C18 modes can not be predicted. For this reason, the difference in energy between the C18 root mode and the maximally overlapping PSTA was calculated, modes that are considered to have not been affected were those in the region of $\omega_{PSTA} = \omega_{C18} \pm 10$\wn. Most modes were within
this region; those that weren't tended to have a significantly different energy. The packing structure had an effect, with PSI being the most unaffected, with 55 modes outside of this energy region (out of 162 modes). PSII, PSIII and PSIV were more affected by the monolayer, with 90, 87 and 71 modes exhibiting a large change
in energy. This effect is larger than that caused by the surface.

The mean characters were compared between the affected and non-affected modes overall (as done for the surface). Those that had been highly affected by the inclusion of monolayer effects were more local, less global, with a higher H character and lower C and S characters. This is explicable because any intrachain interactions would be strongest between the hydrogens of neighbouring chains, as they are the closest approaching parts.

\clearpage

\section{Conclusions}

We have demonstrated that  geometry has a non-negligible effect on 
the dynamics of a ``single-molecule" embedded in a SAM.  These effects can be computed through normal mode analysis, with a simple force-field. Character description of those modes offers a useful qualitative tool in distinguishing the role of each component on the total system, and we expect that these characters will be useful to analyse other systems modelled by normal mode analysius, such as protein folding~\cite{262}. This approach offers a quick way of including geometrical confinement effects on generated conformers including of long chains, which can then be used as a starting point for a more thorough quantum mechanical analysis.

Consideration of substrate and monolayer make-up is necessary for designing SAMs with particular dynamical and structural properties as required for nanotechnological uses. We have shown that the 
gold surface raises the energy of the lowest energy modes, especially those which exhibit highly global longitudinal motion and affect the S adlayer breathing modes. A comparison of the SAM with and without the surface can identify which type of modes are the most sensitive to the surface and thus would be the most affected by a change in surface substrate.

The surrounding monolayer is not an inert substrate. We have demonstrated that minimizing a molecule \textit{in vacuo} will place some modes much higher in energy than they would be in the actual confined system. This has obvious effects for electronic transmission simulations. 
%Future work may cover modeling a different
%molecule inserted into the SAM to confirm that an arbitrary
%molecule will experience geometric confinement in a similar way. 

Regarding the relative strengths of the surface and surrounding monolayer, overall the monolayer had a larger effect on the vibrational frequencies of an inserted molecule and this effect was spread over the whole range of vibrations. However, the surface effect, although smaller, is significant at low energy frequencies and needs to be taken into account if these vibrations are of interest.

This work demonstrates the crucial need to take geometrical confinement into account when using molecular mechanics techniques to find conformers for electrical studies at a higher level of theory. This work concentrated on the simple case of a thioalkane inserted into a thioalkane monolayer, the next step is to repeat this type of analysis with a different molecule.

This research has been supported by EPSRC on grant: GR/S22806/01.

%\section*{References}

\bibliography{lit1}

\end{document}